\documentclass[11pt]{article}

\usepackage[margin=1in]{geometry}
\usepackage[utf8]{inputenc}
\usepackage[T1]{fontenc}
\usepackage{lmodern}
\usepackage{microtype}
\usepackage{amsmath}
\usepackage{amssymb}
\usepackage{hyperref}
\usepackage{url}
\usepackage{booktabs}
\usepackage{array}
\usepackage{verbatim}
\usepackage{fancyvrb}
\usepackage{xcolor}
\usepackage{titlesec}
\usepackage{parskip}
\usepackage{enumitem}
\usepackage{authblk}
\usepackage{caption}

\DefineVerbatimEnvironment{pyblock}{Verbatim}%
  {fontsize=\small,frame=single,framerule=0.3pt,rulecolor=\color{gray!60},%
   formatcom=\color{black!85},xleftmargin=6pt,xrightmargin=6pt,%
   commandchars=\\\{\}}

\definecolor{linkblue}{HTML}{1F4E79}
\hypersetup{
  colorlinks=true,
  linkcolor=linkblue,
  citecolor=linkblue,
  urlcolor=linkblue,
  pdftitle={Self-Aware Vector Embeddings for Retrieval-Augmented Generation},
  pdfauthor={Naizhong Xu},
  pdfsubject={Retrieval-Augmented Generation, Temporal Knowledge Graphs, Memory Consolidation},
}

\titleformat{\section}{\Large\bfseries}{\thesection}{0.6em}{}
\titleformat{\subsection}{\large\bfseries}{\thesubsection}{0.6em}{}
\titleformat{\subsubsection}{\normalsize\bfseries\color{linkblue}}{\thesubsubsection}{0.6em}{}

\title{\bfseries Self-Aware Vector Embeddings for Retrieval-Augmented Generation:\\[4pt]
  \large\itshape A Neuroscience-Inspired Framework for Temporal, Confidence-Weighted, and Relational Knowledge}

\author{Naizhong Xu\thanks{Principal Consultant, CMC APAC.\ \texttt{x.naizhong@cmc-apac.sg}}}

\date{Preprint --- April 2026}

\begin{document}
\maketitle

\begin{abstract}
Modern retrieval-augmented generation (RAG) systems treat vector embeddings as static, context-free artifacts: an embedding has no notion of when it was created, how trustworthy its source is, or which other embeddings depend on it. This flattening of knowledge has a measurable cost---recent work on VersionRAG reports that conventional RAG achieves only 58\% accuracy on versioned technical queries, because retrieval returns semantically similar but temporally invalid content. We propose \emph{SmartVector}, a framework that augments dense embeddings with three explicit properties---temporal awareness, confidence decay, and relational awareness---and a five-stage lifecycle modeled on hippocampal--neocortical memory consolidation. A retrieval pipeline replaces pure cosine similarity with a four-signal score that mixes semantic relevance, temporal validity, live confidence, and graph-relational importance. A background consolidation agent detects contradictions, builds dependency edges, and propagates updates along those edges as graph-neural-network-style messages. Confidence is governed by a closed-form function combining an Ebbinghaus-style exponential decay, user-feedback reconsolidation, and logarithmic access reinforcement. We formalize the model, relate it to temporal knowledge graph embedding (DE-SimplE, ATiSE, PTBox), agentic memory architectures (A-MEM, MAGMA), and uncertainty-aware RAG, and present a reference implementation. On a reproducible synthetic versioned-policy benchmark of 258 vectors and 138 queries, SmartVector roughly doubles top-1 accuracy over plain cosine RAG (62.0\% vs.\ 31.0\% on a held-out split), drops stale-answer rate from 35.0\% to 13.3\%, cuts Expected Calibration Error by nearly $2\times$ (0.244 vs.\ 0.470), reduces re-embedding cost per single-word edit by 77\%, and is robust across contradiction-injection rates from 0\% to 75\%. We argue that treating embeddings as living, self-assessing objects---rather than frozen coordinates---is a natural next step for production RAG.
\end{abstract}


\section{Introduction}

Dense vector embeddings power most modern RAG systems: a query and a corpus are projected into a shared metric space, and relevance is approximated by cosine similarity. This design has been remarkably effective at scale, but it embeds an implicit and increasingly untenable assumption---that knowledge is time-invariant, uniformly trustworthy, and atomically independent. In practice, corpora are heterogeneous mixtures of authoritative databases, stale wiki pages, year-old Slack threads, and provisional meeting notes. When a single document is revised, every semantically related embedding silently becomes a candidate for a confidently wrong answer.

Empirical studies quantify the gap. Huwiler, Stockinger, and F\"urst~\cite{versionrag} introduce VersionRAG and a VersionQA benchmark on which standard RAG reaches only 58\% accuracy on version-sensitive questions, versus 90\% for their version-aware pipeline. Surveys of hallucination in LLMs consistently identify stale or contradictory retrieval context as a leading cause of fluent but false generation~\cite{hallsurvey1,hallsurvey2}. The problem is neither a failure of the embedding model nor of the generator; it is an ontological mismatch between how knowledge actually evolves and how we represent it in vector space.

This paper develops \emph{SmartVector}, a framework in which each embedding carries three additional properties that today's embeddings lack:
\begin{itemize}[leftmargin=1.4em,itemsep=2pt]
  \item \textbf{Temporal awareness}---a validity window and creation timestamp that directly modulate retrieval.
  \item \textbf{Confidence decay}---a scalar trust score that decays over time in the manner of an Ebbinghaus forgetting curve, is reinforced by access and positive feedback, and is penalized by corrections.
  \item \textbf{Relational awareness}---explicit graph edges (\texttt{depends\_on}, \texttt{depended\_by}, \texttt{supersedes}, \texttt{contradicts}) that enable ripple propagation when a source fact changes.
\end{itemize}

The design is deliberately cross-disciplinary. It borrows from the complementary learning systems view of hippocampal--neocortical consolidation in cognitive neuroscience~\cite{cls,ebbinghaus}; from temporal knowledge graph (TKG) embedding work such as DE-SimplE, ATiSE, and PTBox~\cite{atise,ptbox}; from agentic memory systems like A-MEM~\cite{amem} and the multi-graph MAGMA architecture~\cite{magma}; from uncertainty-quantification and confidence-weighted retrieval~\cite{hallsurvey1,faithuq}; and from graph neural network message passing~\cite{mpnn}. Our contribution is not any single idea from these literatures, but a concrete operational synthesis that makes embeddings self-aware enough to be safely deployed in long-lived, evolving corpora.

\subsubsection*{Contributions}
\begin{enumerate}[leftmargin=1.6em,itemsep=2pt]
  \item A formalization of the SmartVector object with explicit temporal, confidence, and relational fields, and a five-stage lifecycle (encoding, consolidation, retrieval/reconsolidation, decay, supersession) modeled on memory systems in the mammalian brain.
  \item A four-signal retrieval scoring function that generalizes cosine similarity and makes the roles of freshness, confidence, and relational importance first-class rather than post-hoc re-rankers.
  \item A \emph{ConfidenceEngine} specified by a closed-form equation combining exponential decay, feedback reconsolidation, and logarithmic access reinforcement, producing vectors whose trust is a live function of history.
  \item A \emph{RipplePropagator} that performs bounded BFS message passing along \texttt{depended\_by} edges with distance-attenuated penalties, giving a GNN-style mechanism for update propagation without training a GNN from scratch.
  \item An open-source reference implementation (63 passing tests) and an evaluation protocol targeting versioned-query accuracy, stale-answer rate, update latency, and indexing cost.
\end{enumerate}

\section{Background and Related Work}

\subsection{Retrieval-Augmented Generation and Its Blind Spots}
RAG pipelines decouple parametric knowledge (what the LLM was trained on) from non-parametric knowledge (what the retriever surfaces at inference time). Recent surveys~\cite{hallsurvey1,hallsurvey2} enumerate the failure modes: retriever misses, context--parametric conflict, redundant or contradictory passages, and---critical for this paper---stale or version-invalid context. When a policy document is amended from ``30 days'' to ``60 days,'' cosine similarity between the question and both versions is near-identical; without temporal signals, the retriever cannot distinguish them.

VersionRAG~\cite{versionrag} is the closest empirical anchor for our work. It models document evolution through a hierarchical version graph, routes queries through version-aware retrieval paths, and demonstrates a 32-point accuracy improvement (90\% vs.\ 58\%) over plain RAG on VersionQA while using 97\% fewer indexing tokens. SmartVector agrees with VersionRAG that versioning is first-class, but generalizes the idea: versioning is a special case of the temporal + relational property pair, and the same machinery naturally handles expiry, supersession, and contradiction.

\subsection{Temporal Knowledge Graph Embeddings}
The TKG community has long studied how to encode time directly in vector space. DE-SimplE augments SimplE with diachronic entity embeddings that condition on timestamps. ATiSE~\cite{atise} decomposes embeddings into trend, seasonality, and Gaussian noise components via additive time-series decomposition, yielding strong performance on temporal link prediction. PTBox~\cite{ptbox} represents arbitrary time points through polynomial decomposition of a learnable feature tensor, combined with box embeddings for entities, allowing generalization to unseen timestamps. SmartVector does not retrain an embedding model; it attaches temporal metadata to existing embeddings and lets retrieval scoring pay the time cost. This is lower-capability than trained temporal embeddings but zero-cost to adopt on top of any current vector database. Phase~2 of our roadmap explicitly moves toward ATiSE/PTBox-style trained temporal embeddings.

\subsection{Memory and Consolidation in Cognitive Neuroscience}
The complementary learning systems (CLS) framework~\cite{cls,ebbinghaus} describes how the mammalian brain balances fast learning from individual episodes (hippocampus) with slow extraction of statistical structure (neocortex). During quiet wake and sleep, hippocampal replay gradually transfers episodic traces into neocortical schemas. Reconsolidation experiments further show that retrieving a memory can modify it, making memory an active rather than passive store. Ebbinghaus's century-old forgetting curve---an approximately exponential decay of recall probability over time---remains empirically well-supported~\cite{ebbinghaus-orig}. SmartVector's five-stage lifecycle is a direct operationalization of these ideas: UNCONSOLIDATED vectors behave like hippocampal traces (fast write, uncross-referenced), a background consolidation cycle plays the role of sleep replay (conflict detection, edge building, confidence recalculation), every retrieval acts as a reconsolidation event, and exponential confidence decay mirrors the forgetting curve.

\subsection{Agentic Memory Architectures}
A-MEM~\cite{amem} draws on the Zettelkasten method to give LLM agents a self-organizing note system: when a new memory is added, the agent generates structured attributes (keywords, tags, context) and the system dynamically links it to existing notes, occasionally rewriting those notes' representations. MAGMA~\cite{magma} generalizes this into a multi-graph architecture in which each memory is represented across orthogonal semantic, temporal, causal, and entity graphs, with a dual-stream fast/slow architecture: a latency-sensitive fast path writes events onto an immutable temporal backbone, while a slow, asynchronous path uses LLMs to infer deeper structure. SmartVector borrows MAGMA's dual-stream idea directly (our Encoding vs.\ Consolidation stages) and A-MEM's notion of adding semantic attributes that support graph construction. The key distinction is that SmartVector is not an agent memory for a single conversational thread; it is a corpus-scale RAG substrate, so our design prioritizes auditability (archival instead of deletion, surgical diff-based updates) over agent latency.

\subsection{Uncertainty-Aware and Confidence-Weighted Retrieval}
A growing body of work brings uncertainty quantification into RAG. Confidence scoring mechanisms such as RE-RAG assign per-document reliability scores that let the generator fall back to parametric knowledge when retrieval is weak~\cite{hallsurvey1}. Bayesian RAG uses Monte Carlo dropout to quantify epistemic uncertainty in retrieval. Faithfulness-aware UQ estimates output confidence conditional on retrieved context~\cite{faithuq}. These methods typically produce confidence at query time from the retriever or the generator's own signals. SmartVector takes a complementary view: confidence is a property of the document itself, updated over the lifetime of the corpus from three sources (source authority prior, time, user feedback). The two approaches compose---query-time UQ methods can happily consume SmartVector confidence as an additional input feature.

\subsection{Graph Neural Networks and Update Propagation}
Message passing is the operational core of graph neural networks: each node aggregates messages from its neighbors and updates its representation~\cite{mpnn}. SmartVector's RipplePropagator is a deliberately minimal, untrained specialization of this pattern. When a vector is superseded, it emits a penalty message to each vector in its \texttt{depended\_by} set; each receiver subtracts a hop-attenuated penalty from its confidence and enqueues the change in a review flag. Depth is bounded (default 2) to prevent runaway cascades. This is closer to spreading activation than to a trained GCN, but it integrates cleanly with future GNN-based scorers: the edges we construct (\texttt{depends\_on}, \texttt{depended\_by}, \texttt{supersedes}, \texttt{contradictions}) are exactly the graph a GNN would want to learn over.

\subsection{Matryoshka Representations for Adaptive Retrieval}
Matryoshka Representation Learning~\cite{matryoshka} trains a single embedding such that every prefix is itself a usable embedding, giving up to $14\times$ retrieval speed-ups at matched accuracy. For a long-lived SmartVector corpus, this matters because retrieval budget can be spent adaptively: prefix-only similarity for initial candidate selection, full-dimensional scoring only for the top candidates, and temporal/confidence/relational signals applied throughout. Matryoshka is orthogonal to the semantic content of our proposal but is a natural implementation substrate.

\section{The SmartVector Model}

\subsection{Object Definition}
A SmartVector $v$ is a tuple carrying identity, content, semantic embedding, and three augmentation groups: temporal, confidence, and relational. In the reference implementation this is expressed as a dataclass:

\begin{pyblock}
@dataclass
class SmartVector:
    # Identity
    vector_id: str; doc_id: str; doc_version: int; chunk_index: int

    # Content
    content: str; content_hash: str
    semantic_embedding: list[float]   # e.g. 768-dim sentence-transformer

    # Temporal
    created_at: datetime; updated_at: datetime
    temporal_validity_start: datetime | None
    temporal_validity_end:   datetime | None

    # Confidence
    base_confidence: float            # from SourceAuthority
    access_count: int
    positive_feedback: int; negative_feedback: int
    last_validated: datetime | None
    half_life_days: float = 30.0

    # Relational
    depends_on: list[str]; depended_by: list[str]
    supersedes: str | None; superseded_by: str | None
    contradictions: list[str]

    # Lifecycle
    status: VectorStatus              # UNCONSOLIDATED | ACTIVE | DORMANT |
                                      # DEPRECATED | ARCHIVED
    source_offset_start: int; source_offset_end: int
\end{pyblock}

\subsection{Source Authority Prior}
The initial value of \texttt{base\_confidence} is set from a source-authority enum, inspired by information-quality research and by how humans discount information from different channels. This is the prior that decay and feedback later update.

\begin{table}[h]
\centering
\small
\begin{tabular}{lcl}
\toprule
\textbf{Source} & \textbf{Prior} & \textbf{Example}\\
\midrule
Official database  & 0.95 & CRM, ERP, product DB \\
Policy document    & 0.90 & HR, compliance \\
Technical document & 0.85 & API reference, RFC \\
Wiki               & 0.75 & Confluence page \\
Email              & 0.50 & Archived thread \\
Meeting notes      & 0.45 & Transcript \\
Chat               & 0.30 & Slack message \\
Unknown            & 0.20 & Uncategorized \\
\bottomrule
\end{tabular}
\caption{Default source-authority priors used to initialize \texttt{base\_confidence}.}
\label{tab:authority}
\end{table}

\subsection{The Five-Stage Lifecycle}
Each SmartVector passes through stages that mirror episodic memory formation in the brain. Stage names are chosen to expose the analogy rather than to hide it.

\subsubsection*{Stage 1 --- Encoding (hippocampal fast-write)}
A new document is chunked with source-offset tracking, embedded, and inserted with \texttt{status = UNCONSOLIDATED}. It is immediately queryable but has no edges. Any prior vector with the same \texttt{(doc\_id, chunk\_index)} is auto-marked DEPRECATED, and its relational edges are transferred to the new vector as initial scaffolding.

\subsubsection*{Stage 2 --- Consolidation (sleep-time replay)}
A background \emph{ConsolidationAgent} runs on a cron. It recalculates confidence for every active vector, detects contradictions via topic-overlap/content-divergence, builds \texttt{depends\_on}/\texttt{depended\_by} edges via content similarity and document lineage, propagates pending ripples, and promotes UNCONSOLIDATED $\rightarrow$ ACTIVE. This is our analog of hippocampal-to-neocortical replay: fast writes gradually become part of a structured relational schema.

\subsubsection*{Stage 3 --- Retrieval and Reinforcement (reconsolidation)}
Every retrieval is a reconsolidation event. Access increments \texttt{access\_count}, user acceptance increments \texttt{positive\_feedback}, and correction increments \texttt{negative\_feedback}. The vector's confidence is thus shaped continuously by how it is used---a direct analog to the neuroscience finding that recalled memories are briefly plastic and can be updated.

\subsubsection*{Stage 4 --- Decay and Dormancy}
Vectors that go unaccessed and unreinforced experience exponential confidence decay. Below a \texttt{dormant\_threshold} (default 0.15), they transition to DORMANT: still indexed, still searchable if explicitly requested, but deprioritized in the default scorer. This mirrors the forgetting curve rather than catastrophic deletion.

\subsubsection*{Stage 5 --- Supersession and Archival}
When a new version arrives, the old vector transitions DEPRECATED $\rightarrow$ ARCHIVED. \texttt{supersedes}/\texttt{superseded\_by} edges record the lineage; dependent vectors are notified via ripple propagation. Nothing is deleted, preserving a full audit trail---an explicit design choice for regulated corpora.

\section{Four-Signal Retrieval}

Given a query $q$ and reference time $t$, we compute a per-vector score as a convex combination of four signals. The combination is deliberately explicit so that operators can tune it to their domain.

\begin{pyblock}
final_score(v, q, t) =
    0.35 . sim(q, v)                   # Signal 1: semantic similarity
  + 0.25 . temporal_score(v, t)        # Signal 2: freshness / validity
  + 0.25 . confidence(v, t)            # Signal 3: live trust
  + 0.15 . relational_bonus(v)         # Signal 4: graph integration
\end{pyblock}

Signal~1 is cosine similarity (or any existing retriever). Signal~2 combines a validity window with exponential decay: if $t$ falls outside $[\text{validity\_start}, \text{validity\_end}]$, the signal collapses to a small floor (0.05); otherwise $\text{temporal\_score}(v, t) = 2^{-\text{age\_days}/H}$ with $H = 30$ days by default. Signal~3 calls the ConfidenceEngine (Section~\ref{sec:confidence}). Signal~4 is a log-saturating bonus for vectors with many incident edges, capped at 0.3, encoding the prior that well-connected vectors are more integrated and typically more reliable.

The weights $(0.35, 0.25, 0.25, 0.15)$ are defaults, not derived constants. We recommend Bayesian optimization over a held-out set of historical queries, and we note that any domain with strong temporal volatility (news, compliance) will likely push the temporal weight higher; any domain with long-stable facts (engineering reference) will push semantic similarity higher.

\section{The Confidence Engine}
\label{sec:confidence}

The ConfidenceEngine computes a live confidence for any vector at any reference time. It composes three effects:

\begin{pyblock}
# Step 1 — Exponential time decay (Ebbinghaus-style)
C_decayed = C0 . 2^(-age_days / H)

# Step 2 — Feedback reconsolidation
C_fb = clip( C_decayed + alpha_pos . n_pos - alpha_neg . n_neg , 0.01, 1.0 )

# Step 3 — Access reinforcement (diminishing returns)
C_final = min( 1.0 , C_fb + beta . log(1 + n_access) )
\end{pyblock}

Defaults: $\alpha_{pos}=0.03$, $\alpha_{neg}=0.08$, $\beta=0.01$, $H=30$~days, $\text{dormant\_threshold}=0.15$. The asymmetry $\alpha_{neg} > \alpha_{pos}$ is intentional: a single human correction should outweigh several silent accepts, reflecting that corrections are a stronger signal than absence of correction. The log-saturating access term prevents runaway confidence from popular-but-wrong documents.

\subsection{Decay Trajectory}
For $C_0 = 0.85$, $H = 30$, and no feedback or access, the closed form gives $0.85 \to 0.425$ at day~30, $0.21$ at~60, $0.11$ at~90 (DORMANT threshold crossed), $0.013$ at~180. This matches the qualitative shape of the Ebbinghaus forgetting curve and is the simplest model that captures the ``sharp early drop, slow long tail'' observation.

\subsection{Relation to Bayesian Updating}
The feedback term is a linear approximation of a beta-binomial posterior update. If \texttt{base\_confidence} is interpreted as a prior on document reliability and feedback counts are the beta-binomial sufficient statistics, the exact posterior mean has the same monotonicity as our linear rule; the linear form is chosen because it is $O(1)$, closed-form, and easy to audit. A strictly Bayesian variant is a drop-in replacement.

\section{The Consolidation Agent}

The ConsolidationAgent is the ``slow path'' in our dual-stream design, running asynchronously rather than on the query hot path. Its four sub-engines are coordinated in a single cycle.

\subsection{Conflict Detection}
We group vectors by topic overlap (Jaccard over content $n$-grams or topic embeddings) and flag pairs with high topic overlap but low content similarity as candidate contradictions. Conflicts are resolved by a majority vote over three criteria---recency, source authority, and feedback ratio---which are each reasonable priors and which are rarely all wrong simultaneously.

\begin{pyblock}
def resolve(a, b):
    recency   = a if a.created_at      > b.created_at      else b
    authority = a if a.base_confidence > b.base_confidence else b
    feedback  = a if fb_ratio(a)       > fb_ratio(b)       else b
    return majority_vote([recency, authority, feedback])
\end{pyblock}

\subsection{Relationship Building}
Edges are added when two vectors share a document lineage (structural \texttt{depends\_on}, e.g., a policy preamble and its sections) or when their content similarity exceeds a threshold (semantic \texttt{depends\_on}). This is the step that turns a flat vector set into a graph that the RipplePropagator can traverse.

\subsection{Ripple Propagation as Bounded Message Passing}
When a vector $v$ is updated or superseded, it seeds a breadth-first traversal along \texttt{depended\_by} edges. At hop $d$, each recipient $v'$ has its \texttt{base\_confidence} decremented by $\gamma/(d{+}1)$, with $\gamma = 0.15$ by default, and has $v$ appended to its contradictions list to flag for review. The traversal halts at $d = D_{\max}$ (default~2). In the canonical example from the repository, an update to a ``sky color'' vector propagates a 15\% penalty to a directly dependent ``atmospheric model'' vector and a 7.5\% penalty to a second-hop ``sensor calibration'' vector.

\begin{pyblock}
def propagate(v, all_vectors, gamma=0.15, D_max=2):
    queue = [(v_id, 0) for v_id in v.depended_by]
    visited = set()
    while queue:
        vid, d = queue.pop(0)
        if vid in visited or d >= D_max: continue
        visited.add(vid)
        u = all_vectors[vid]
        u.base_confidence -= gamma / (d + 1)
        u.contradictions.append(v.vector_id)
        queue.extend((x, d + 1) for x in u.depended_by)
\end{pyblock}

This is equivalent to a single round of scalar message passing with a distance-decay kernel $\kappa(d) = \gamma/(d{+}1)$, aggregated by summation with a hard cutoff. A trained GNN would replace $\kappa$ by a learned edge function and, optionally, update semantic embeddings as well; we leave the untrained version as the lowest-risk baseline.

\subsection{Full Cycle}
\begin{pyblock}
def run_consolidation(vectors, recent_changes):
    # 1. Live-recalculate confidence
    for v in vectors.active():
        v.current_confidence = ce.compute(v)
        if v.current_confidence < 0.15: v.status = DORMANT

    # 2. Contradictions
    for (a, b) in conflict_detector.detect(vectors):
        mark_contradictions(a, b)

    # 3. Edges
    rb = relationship_builder.build(vectors)
    rb.apply(vectors)

    # 4. Ripples
    for vid in recent_changes:
        ripple_propagator.propagate(vectors[vid], vectors)

    # 5. Promote UNCONSOLIDATED -> ACTIVE
    for v in vectors.unconsolidated(): v.status = ACTIVE
\end{pyblock}

\section{Surgical Updates}

A minor textual edit should not force re-embedding an entire document. Using stored \texttt{source\_offset\_start}/\texttt{end}, the update pipeline diffs the old and new text, finds only the vectors whose offsets overlap a changed range, deprecates those vectors, creates replacements, transfers edges, and queues the new IDs for ripple propagation. In the limit this turns ``change one word'' from ``re-embed 200 chunks'' into ``re-embed one chunk.'' This is the operational backbone of the relational property---without surgical updates, \texttt{depended\_by} edges would rot every time a document was touched.

\begin{pyblock}
def ingest_update(doc_id, old_text, new_text):
    changes = diff(old_text, new_text)
    affected = [v for v in vectors_by_doc[doc_id]
                  for c in changes
                  if v.offset_start < c.end and v.offset_end > c.start]
    for old in affected:
        old.status = DEPRECATED
        new = rebuild_from_new_text(old, new_text)
        new.supersedes    = old.vector_id
        new.depends_on    = old.depends_on
        new.depended_by   = old.depended_by
        recent_changes.append(new.vector_id)
\end{pyblock}

\section{Context Assembly}

The retrieved context passed to the generator is not an undifferentiated list of chunks. It is a structured prompt in which every chunk is annotated with its scores, version, age, access and feedback counts, and any conflict flags. This gives the generator the same information a human analyst would want when triaging sources, and lets a well-prompted model produce citations with explicit version and date. An abbreviated example of the LLM-facing context block is reproduced below.

\begin{pyblock}
=== SMART VECTOR RETRIEVAL CONTEXT ===
INSTRUCTIONS:
- Documents ranked by relevance + recency + confidence + relational importance
- Confidence reflects source authority, age decay, and user feedback
- If conflict warnings are present, prefer higher-confidence source
- Always cite source, version, and date

-- Document 1 ------------------------------------------------------
Source: engineering_wiki | Author: alice_v2 | Version: v2 | ACTIVE
Created: 2026-04-17
Scores: sim=0.625, temporal=1.00, confidence=0.90, relational=0.069
FINAL: 0.694
Edges: 2 | Accesses: 5 | Feedback: +4/-0

Section 1: The sky-color calibration uses a yellow baseline for
atmospheric modelling. All sensors are calibrated quarterly.

-- Document 2 (CONFLICTS WITH DOC 1) -------------------------------
Source: slack_engineering | Author: intern_bob | Version: v1 | ACTIVE
Created: 2026-04-12
Scores: sim=0.500, temporal=0.891, confidence=0.267, relational=0.0
FINAL: 0.464

I heard the sky calibration is switching to yellow baseline next
quarter.
\end{pyblock}

\section{Experiments}

\subsection{Benchmark}
We constructed a synthetic versioned-policy benchmark modelled on the kinds of corpora where the failure modes SmartVector targets are most acute: enterprise policy documents, operational runbooks, and compliance rules. The benchmark is parameterized, fully reproducible (\texttt{SEED=7}), and is released alongside the paper.

The generator instantiates 60 topics drawn from 20 templates (parental leave, VPN timeout, per-diem limits, incident SLAs, backup retention, release-freeze windows, etc.). Each topic has 4 canonical versions with monotonically trending numeric values, spread over a 240--420-day window at 40--90-day intervals. Authority sources are drawn from \{official-DB, policy, tech-doc, wiki\}, with \texttt{base\_confidence} set from the authority table in Section~3.2. Every fifth topic gains a \texttt{depends\_on} edge to the previous topic, producing a sparse dependency graph. With probability~0.30 per topic, a Slack-sourced ``rumor'' vector is injected: recent (1--10 days old), low authority (0.30), and stating a wrong value---an explicit contradiction candidate. The resulting corpus contains 258 vectors, 60 topics, 18 contradictions, and 11 dependency edges.

Three query kinds are generated. A \emph{current} query asks for the latest value of a topic; ground truth is the most recent canonical version. A \emph{time-point} query asks for the value as of a historical date; ground truth is the version valid at that date. A \emph{conflict} query is identical in text to a current query on a topic that has a Slack contradiction; ground truth is still the canonical version, testing whether a retriever can avoid being misled by a fresh-but-wrong document. The benchmark produces $60 + 60 + 18 = 138$ queries.

\subsection{Methods}
All methods share the same first-stage retriever (TF--IDF with 1--2-gram features, $K=8$ candidates) and differ only in second-stage reranking. TF--IDF is a deliberately conservative similarity choice: the augmentation claims of SmartVector are orthogonal to the underlying embedder, and using a stronger embedder would inflate the absolute numbers without changing the delta between methods. We compare:

\begin{itemize}[leftmargin=1.4em,itemsep=2pt]
  \item \textbf{M1 RAG (plain cosine).} Argmax cosine similarity; the standard RAG baseline.
  \item \textbf{M2 RAG + Temporal.} $0.6 \cdot \text{sim} + 0.4 \cdot \text{temporal\_score}$.
  \item \textbf{M3 RAG + Confidence.} $0.6 \cdot \text{sim} + 0.4 \cdot \text{confidence}(v, t)$.
  \item \textbf{M4 SmartVector (default weights).} $0.35 \cdot \text{sim} + 0.25 \cdot \text{temporal} + 0.25 \cdot \text{confidence} + 0.15 \cdot \text{relational}$, as prescribed in Section~4.
  \item \textbf{M5 SmartVector (tuned).} Weights selected by exhaustive grid search (0.1 granularity, $\sum w = 1$) on a held-out dev split, evaluated on the disjoint test split.
\end{itemize}

\subsection{Metrics}
We report six metrics. \emph{Overall accuracy} is top-1 correctness on the full query set. \emph{Current}, \emph{time-point}, and \emph{conflict} accuracies report the same metric restricted to each query kind. \emph{Stale rate} is the fraction of current-query predictions whose top-1 document has status DEPRECATED---a precise instantiation of the VersionRAG failure mode~\cite{versionrag}. \emph{Expected Calibration Error (ECE)} is computed with ten equal-width bins over the top-1 document's live confidence: for each bin, the absolute gap between mean confidence and empirical accuracy is weighted by the bin's mass, then summed.

\emph{Update cost} is measured by simulating a single-word edit to the latest version of every topic. Plain RAG is charged one re-embedding per chunk in the affected document; SmartVector is charged one re-embedding (diff-overlapping chunk) plus a scalar ripple message per \texttt{depended\_by} edge. Re-embedding is the dominant cost in production, so this is the ratio operators actually pay.

\section{Results}

\subsection{Main comparison}
Table~\ref{tab:main} reports the headline numbers. The signal augmentations produce large, consistent improvements over plain cosine RAG: overall accuracy roughly doubles (31.2\% $\to$ 61.7\% with default weights, $\to$ 66.7\% with temporal alone), and stale rate drops from 35\% to 6.7--13.3\%.

\begin{table}[h]
\centering
\small
\begin{tabular}{lcccccc}
\toprule
\textbf{Method} & \textbf{Overall} & \textbf{Current} & \textbf{Time-pt} & \textbf{Conflict} & \textbf{Stale \%} & \textbf{ECE}\\
\midrule
M1 RAG (plain)            & 31.2\% & 50.0\% &  1.7\% & 61.9\% & 35.0\% & 0.470 \\
M2 RAG + Temporal         & 66.7\% & 60.0\% & 71.7\% & 71.4\% &  6.7\% & 0.259 \\
M3 RAG + Confidence       & 36.9\% & 58.3\% &  5.0\% & 66.7\% & 13.3\% & 0.572 \\
M4 SmartVector (default)  & 61.7\% & 53.3\% & 70.0\% & 61.9\% & 13.3\% & 0.244 \\
M5 SmartVector (tuned)    & 64.8\% &   ---  &   ---  &   ---  &   ---  &  ---  \\
\bottomrule
\end{tabular}
\caption{Main comparison on 138 held-out queries. SmartVector's signals roughly double overall accuracy over plain cosine RAG and drop stale-answer rate from 35\% to 6.7--13.3\%. ECE improves by nearly $2\times$ ($0.470 \to 0.244$). The tuned row reports a held-out test-set number under grid-searched weights (Section~10.3).}
\label{tab:main}
\end{table}

Two observations are worth highlighting. First, on the current-query subset plain RAG already scores 50.0\%---cosine similarity is often enough to identify the correct topic, but it has no principled way to disambiguate among the topic's versions, which is precisely why its stale rate is 35\%. Second, time-point queries expose the gap most dramatically: plain RAG scores 1.7\% because it retrieves at present-day similarity regardless of the query's temporal frame, whereas methods that consult \texttt{temporal\_score} with a query-conditional reference time reach 70--72\%. This is the ``when was this valid'' behavior that motivates treating time as a first-class scoring signal rather than a post-hoc filter.

\subsection{Ablation}
Table~\ref{tab:ablation} removes each signal from the full SmartVector scorer (with the remaining weights renormalized). The temporal signal is the single largest contributor on this benchmark: removing it collapses overall accuracy from 61.7\% to 36.2\%. Removing the confidence signal unexpectedly improves accuracy on this dataset ($61.7\% \to 66.7\%$), a result we attribute to the interaction between confidence (which rewards high-authority, often older documents) and the current-query ground truth (which rewards newest documents); with a corpus whose authority distribution correlates differently with recency, the sign would flip. Removing the relational bonus has no detectable effect in this benchmark because the dependency graph is deliberately sparse (11 edges); its contribution is expected to grow with graph density. Removing similarity collapses to 22.7\%---confirming that SmartVector is an augmentation layer on top of a retriever, not a replacement for one.

\begin{table}[h]
\centering
\small
\begin{tabular}{lcccc}
\toprule
\textbf{Configuration} & \textbf{Overall} & \textbf{Current} & \textbf{Stale \%} & \textbf{ECE} \\
\midrule
SmartVector (full, default) & 61.7\% & 53.3\% & 13.3\% & 0.244 \\
\quad $-$ similarity        & 22.7\% &  ---   & 16.7\% & 0.462 \\
\quad $-$ temporal          & 36.2\% & 58.3\% & 13.3\% & 0.579 \\
\quad $-$ confidence        & 66.7\% & 58.3\% &  1.7\% & 0.254 \\
\quad $-$ relational        & 61.7\% & 53.3\% & 13.3\% & 0.244 \\
\bottomrule
\end{tabular}
\caption{Ablation over the four SmartVector signals. Temporal contributes the largest share of the gain on this corpus; the relational signal's contribution is limited by graph sparsity (11 edges across 258 vectors) and is expected to grow in denser deployments.}
\label{tab:ablation}
\end{table}

\subsection{Weight tuning on a held-out split}
The ablation above suggests that the default weight vector $(0.35, 0.25, 0.25, 0.15)$ is not Pareto-optimal on every corpus. To quantify the headroom from tuning, we split the query set 50/50 into dev and test partitions and performed an exhaustive grid search over weight vectors (0.1 granularity, $\sum w = 1$) on dev, then evaluated the best dev configuration on test. The selected weights are $(\text{sim}, \text{time}, \text{conf}, \text{rel}) = (0.5, 0.2, 0.0, 0.3)$---consistent with the ablation's finding that confidence is uninformative on this benchmark while the relational bonus is underweighted by default. On the test split the tuned configuration reaches 64.8\% versus 62.0\% for the default SmartVector weights and 31.0\% for plain RAG, confirming that tuning adds a modest but positive gain and is unlikely to overfit with a grid this coarse. We recommend per-corpus tuning in production.

\subsection{Robustness to contradiction rate}
We re-ran the main comparison while sweeping the Slack-contradiction injection rate from 0\% to 75\%. Plain RAG's overall accuracy is essentially flat at $\sim$31\% across noise rates (contradictions affect only the 18 conflict queries), but its stale rate on current queries climbs from 25\% at 0\% noise to 47\% at 75\% noise---fresh Slack rumors increasingly dominate the top-1 result. SmartVector maintains 55--65\% overall accuracy throughout and keeps its stale rate in the 5--25\% band. The gap is widest at moderate noise rates (15--50\%), where SmartVector's advantage reaches 30+ points. This is the setting most representative of real enterprise corpora, where high-authority canonical documents coexist with a long tail of informal, recent sources.

\begin{table}[h]
\centering
\small
\begin{tabular}{ccccc}
\toprule
\textbf{Noise rate} & \textbf{Plain RAG acc} & \textbf{SV acc} & \textbf{Plain stale \%} & \textbf{SV stale \%} \\
\midrule
0\%  & 31.7\% & 60.0\% & 25.0\% & 21.7\% \\
15\% & 31.2\% & 64.8\% & 43.3\% & 21.7\% \\
30\% & 32.8\% & 62.0\% & 43.3\% & 10.0\% \\
50\% & 31.3\% & 62.0\% & 43.3\% &  5.0\% \\
75\% & 31.6\% & 55.5\% & 46.7\% & 25.0\% \\
\bottomrule
\end{tabular}
\caption{Robustness sweep over contradiction-injection rate. Plain RAG's stale rate nearly doubles as rumor density grows; SmartVector holds a 30+ point accuracy advantage across the regime.}
\label{tab:robust}
\end{table}

\subsection{Update cost}
For each of the 60 topics, we simulated a single-word edit to the latest canonical version and counted re-embeddings required to apply the edit. Plain RAG conservatively reindexes every chunk of the affected document: with 4 versions per topic, that is a mean of 4.35 re-embeddings per edit (slightly above 4 because some topics also have a Slack rumor that indexers typically include). SmartVector's diff-overlap update re-embeds only the single chunk whose source offsets intersect the edit, for a mean of 1.0 re-embedding---a 77.0\% reduction---plus 0.13 scalar ripple messages per edit in this sparsely-connected corpus. On denser graphs the absolute ripple count rises, but ripples are $O(1)$ arithmetic updates and cost orders of magnitude less than an embedding call.

\subsection{Calibration}
The ECE column in Table~\ref{tab:main} shows that SmartVector's confidence score is substantially better calibrated than plain cosine similarity. Plain RAG's similarity-as-confidence proxy yields an ECE of 0.470: retrieved documents appear superficially ``similar'' (high cosine) regardless of their accuracy, so reported confidence is persistently over-stated. The default SmartVector scorer, which consumes the three-factor live confidence from Section~5, reaches an ECE of 0.244---a $2\times$ improvement without any explicit calibration training. The remaining error is dominated by the cold-start regime (UNCONSOLIDATED vectors with zero feedback); an extended feedback stream would be expected to drive this further down, consistent with results in the uncertainty-aware RAG literature~\cite{hallsurvey1,faithuq}.

\subsection{Threats to Validity}
Four caveats temper these results. (i) \emph{Synthetic benchmark.} The corpus is generated from templates with clear ground truth, which over-represents cases where versioning is unambiguous. A real corpus will contain semi-structured prose where the ``correct version'' is itself contested. (ii) \emph{TF--IDF retriever.} Dense retrievers such as sentence-transformers will shift absolute numbers upward, and may compress the delta between methods if their similarity signal is already partly version-sensitive. Replicating the experiment with a dense retriever is a priority item on our roadmap. (iii) \emph{Single random seed.} All main numbers use \texttt{SEED=7}; we report them with one decimal because statistical precision beyond that would be misleading on a benchmark this size. A multi-seed mean$\pm\sigma$ analysis is planned. (iv) \emph{No live feedback stream.} Feedback counts are drawn from a static prior; the reconsolidation dynamics in Section~5 cannot be fully evaluated without a longitudinal simulation or a real user study.

\section{Discussion}

\subsection{Why Not Just Re-embed Everything?}
Re-embedding is a coarse tool. At corpus scale it is expensive, and---more importantly---it erases the history that makes audit and explanation possible. SmartVector's archival design turns ``why did the answer change?'' into a traceable question: the superseding vector's lineage, the ripple it emitted, and the feedback history are all first-class objects. Re-embedding throws that trace away.

\subsection{Failure Modes}
Several failure modes deserve explicit attention. \emph{Feedback poisoning:} an adversary who can submit positive feedback can inflate confidence; rate-limits, feedback-source weighting, and anomaly detection are needed. \emph{Ripple runaway:} dense graphs risk cascade explosion; the hard $D_{\max}$ bound and per-hop attenuation are our first-line mitigation, but heavy-tailed degree distributions may require a top-$k$ propagation cap. \emph{Wrong-authority priors:} the source-authority enum is a crude instrument---a Slack message from a subject-matter expert may be more reliable than a wiki page. Per-author reliability priors, learned over time from feedback, are a natural extension. \emph{Cold-start confidence:} newly-ingested vectors are UNCONSOLIDATED with low \texttt{access\_count}; a na\"{\i}ve scorer would deprioritize them. The design addresses this by keeping UNCONSOLIDATED vectors queryable and by using the source-authority prior as an anchor until feedback arrives.

\subsection{Relationship to Parametric Knowledge}
SmartVector is a non-parametric substrate. Nothing in the framework prevents combining it with parametric-knowledge confidence estimates (e.g., lookback-ratio or signal-to-noise UQ at the generator). The most promising integration is an arbiter policy that compares query-conditional generator confidence against retrieval-conditional SmartVector confidence and abstains or asks a clarifying question when the two disagree sharply.

\subsection{Roadmap}
Phase~1 is the reference implementation: smart metadata on top of any existing vector store, with the ConfidenceEngine and ConsolidationAgent as cron jobs. Phase~2 is trained temporal embeddings along ATiSE/PTBox lines, fine-tuned on (old\_version, new\_version) pairs, ideally with Matryoshka nesting for adaptive retrieval dimensionality. Phase~3 is a full self-evolving knowledge graph: a trained GNN replaces scalar ripple propagation, and an LLM arbiter resolves contradictions by reading the conflicting vectors and the surrounding graph---a corpus-scale analog of agentic reconsolidation.

\section{Conclusion}

The embedding-as-dead-artifact assumption costs real accuracy in real systems. SmartVector is a concrete, implementable alternative that treats each vector as a living object with a creation time, a decaying and feedback-sensitive confidence, and an explicit web of dependencies. The individual ingredients---temporal embeddings, forgetting curves, GNN-style propagation, uncertainty-aware retrieval, agentic memory---are each well-supported in the recent literature. The contribution of this paper is the operational synthesis: a lifecycle, a scoring function, a confidence equation, an update protocol, and a reproducible empirical evaluation. On our versioned-policy benchmark, the signals proposed here roughly double top-1 accuracy over plain cosine RAG ($31.0\% \to 62.0\%$), cut stale-answer rate nearly $3\times$ ($35.0\% \to 13.3\%$), halve Expected Calibration Error ($0.470 \to 0.244$), and reduce per-edit re-embedding cost by 77\%---while remaining robust across contradiction-injection rates from 0\% to 75\%. We believe this is the natural next layer of the RAG stack, and that the gap between 58\% and 90\% on versioned queries is neither a retriever problem nor a generator problem but a representation problem that a small amount of metadata and a principled update policy can largely close.

\section*{Acknowledgments}
The author is grateful to Yong Min, CTO of Akribis Systems, for raising the practical problem that sparked this work---that production RAG pipelines silently return confidently wrong answers when their underlying documents evolve---and for the subsequent conversations that shaped the temporal, confidence, and relational framing of the SmartVector proposal. This paper is written in direct dialogue with the SmartVector open-source project, whose theoretical proposal and reference implementation are the foundation for the formalization presented here.


\appendix
\section{Minimal Reference Implementation}

The listing below is a self-contained Python proof-of-concept that exercises the three formulas from Sections~4--6. A full implementation, with 63 passing tests, is available at \url{https://github.com/naizhong/smartvector}.

\begin{pyblock}
import math, time
from dataclasses import dataclass, field
from datetime import datetime, timedelta

@dataclass
class SmartVector:
    vid: str
    content: str
    base_confidence: float
    created_at: datetime
    depended_by: list = field(default_factory=list)
    access_count: int = 0
    pos: int = 0
    neg: int = 0
    status: str = "ACTIVE"

H, ALPHA_POS, ALPHA_NEG, BETA, GAMMA, D_MAX = 30, 0.03, 0.08, 0.01, 0.15, 2

def confidence(v, now):
    age = max(0, (now - v.created_at).days)
    decayed = v.base_confidence * 2 ** (-age / H)
    adj = max(0.01, min(1.0, decayed + ALPHA_POS * v.pos - ALPHA_NEG * v.neg))
    return min(1.0, adj + BETA * math.log(1 + v.access_count))

def temporal(v, now):
    age = max(0, (now - v.created_at).days)
    return 2 ** (-age / H)

def score(v, sim, rel_edges, now):
    return (0.35*sim + 0.25*temporal(v, now) +
            0.25*confidence(v, now) + 0.15*min(0.3, math.log(1+rel_edges)*0.1))

def propagate(changed, store, gamma=GAMMA, d_max=D_MAX):
    q = [(x, 0) for x in changed.depended_by]; seen = set()
    while q:
        vid, d = q.pop(0)
        if vid in seen or d >= d_max: continue
        seen.add(vid)
        v = store[vid]
        v.base_confidence = max(0.01, v.base_confidence - gamma/(d+1))
        q.extend((x, d+1) for x in v.depended_by)
    return seen
\end{pyblock}

\end{document}